\documentclass[a4paper,11pt]{article}
\usepackage{pos}
\usepackage[english]{babel}
\usepackage{csquotes}
\usepackage{scalefnt}
\usepackage{braket}
\usepackage{enumitem}
\usepackage{acronym}
\usepackage{tikz}
\usepackage{xcolor}
\usepackage[capitalise]{cleveref}

\renewcommand{\logo}{\relax}
\notes{\note[]{TTK-23-29  / P3H-23-081 — October 2023}}  
\renewcommand{\hookAfterAbstract}{%
\vspace{2em}
\begin{center}
\includegraphics[width=.3\textwidth]{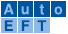}
\end{center}
}

\Crefname{equation}{Eq.}{Eqs.}
\setlist[description]{style=nextline}
\usetikzlibrary{shapes.geometric, arrows}

\newcommand*{\abbrev}[1]{{\scalefont{.9}#1}}
\newcommand*{\lhs}{\abbrev{LHS}}
\newcommand*{\rhs}{\abbrev{RHS}}
\newcommand*{\citere}[1]{Ref.~\cite{#1}}
\newcommand*{\citeres}[1]{Refs.~\cite{#1}}
\newcommand*{\code}[1]{\texttt{#1}}
\newcommand*{\autoeft}{\code{AutoEFT}}
\newcommand*{\python}{\code{Python}}
\newcommand*{\sage}{\code{SageMath}}
\newcommand*{\gepris}[1]{\href{https://gepris.dfg.de/gepris/projekt/#1?language=en}{#1}}

\DeclareMathOperator{\SO}{SO}
\DeclareMathOperator{\U}{U}
\DeclareMathOperator{\SU}{SU}
\DeclareMathOperator{\LG}{\SO^+}

\newcommand{\myacrodef}[3]{\acrodef{#2}{#3}\newcommand{#1}{\ac{#2}}}
\myacrodef{\sm}{SM}{Standard Model}
\myacrodef{\eft}{EFT}{Effective Field Theory}
\acrodefplural{EFT}{Effective Field Theories}
\newcommand{\efts}{\acp{EFT}}
\myacrodef{\smeft}{SMEFT}{Standard Model Effective Field Theory}
\myacrodef{\ibp}{IbP}{Integration by Parts}
\myacrodef{\eom}{EoM}{Equation of Motion}
\acrodefplural{EoM}{Equations of Motion}
\newcommand{\eoms}{\acp{EoM}}
\myacrodef{\rfr}{RFR}{Repeated Field Redundancy}
\acrodefplural{RFR}{Repeated Field Redundancies}
\newcommand{\rfrs}{\acp{RFR}}

\title{AutoEFT\@: Constructing and exploring on-shell bases of effective field theories}

\author*{Magnus~C.~Schaaf}

\affiliation{Institute for Theoretical Particle Physics and Cosmology,\\
RWTH Aachen University, 52056 Aachen, Germany}

\emailAdd{magnus.schaaf@rwth-aachen.de}

\abstract{%
\efts\ provide a framework for capturing the effects of yet unseen heavy degrees of freedom in a model-independent manner.
However, constructing a complete and minimal set of operators, especially at higher mass dimensions, is challenging.
We present \autoeft, an implementation of an algorithm that systematically handles redundancies among operators due to equations of motion, integration-by-parts identities, Fierz identities, and repeated fields.
This algorithm enables the construction of on-shell bases for a broad range of \efts.
Additionally, it facilitates the exploration of various aspects within this field, such as investigating higher mass dimensions or the relationship between different operator bases.
\autoeft\ can be applied to phenomenologically relevant theories like the Standard Model and its extensions, including new light particles or additional symmetry groups.%
}

\FullConference{The European Physical Society Conference on High Energy Physics (EPS-HEP2023)\\
 21-25 August 2023\\  
Hamburg, Germany\\}

\begin{document}
\maketitle
\acresetall%

\section{Introduction}
Since current and potential future colliders have a limited energy reach, the on-shell discovery of new heavy particles beyond the \sm\ may be challenging.
In recent years, there has been a growing interest in searching for new physics in virtual effects.
To correctly associate any deviation to new physics, these effects need to be parameterized systematically.
For this approach, \efts\ have become invaluable.

In this article, we feature \autoeft~\cite{Harlander:2023ozs}, an independent open-source implementation of a systematic algorithm~\cite{Li:2020gnx,Li:2020xlh} developed for the on-shell operator basis construction of \efts.
\autoeft\ is published under the \abbrev{MIT} license~\cite{mit_license} and is implemented in \python~\cite{python}, combined with the open-source mathematics software system \sage~\cite{sagemath}.
It is available for download on the \emph{Python Package Index}~\cite{pypi} and the \emph{conda-forge}~\cite{conda_forge_community_2015_4774216} community channel.
For installation details and the usage of \autoeft, we refer the reader to \citere{Harlander:2023ozs}.

\section{Effective field theories}
In the \emph{bottom-up} formulation of an \eft, the renormalizable Lagrangian $\mathcal{L}^{(\le4)}$, describing a low-energy theory, is extended by effective interactions mediated by some unknown high-energy physics.
These interactions are encoded in so-called \emph{operators} $\mathcal{O}_i^{(d)}$, composed out of the field content of $\mathcal{L}^{(\le4)}$, and $d$ denoting the combined mass dimension of the participating fields.
For consistency with the low-energy theory, the operators must be invariant under the symmetries imposed on $\mathcal{L}^{(\le4)}$.
Depending on the scenario, one may even impose additional symmetries on the operators, for example, some of the accidental or approximate symmetries of $\mathcal{L}^{(\le4)}$.

Since any operator with a mass dimension equal to or less than four is already part of $\mathcal{L}^{(\le4)}$, the effective operators are necessarily to higher mass dimension.
To cancel the surplus mass dimension of the operators $\mathcal{O}_i^{(d)}$ in the Lagrangian, they must be multiplied by an additional mass scale $\Lambda^{4-d}$, where $\Lambda$ is related to heavy new physics.
The effects of these operators contribute with a factor ${(E/\Lambda)}^{d-4}$ to the physical amplitudes, where $E$ denotes the typical energy of the low-energy process.
Assuming the mass scales obey the hierarchy $E\ll\Lambda$, the effects up to order ${(E/\Lambda)}^{\mathcal{N}}$ can be incorporated by
\begin{equation}\label{eq:eft}
\mathcal{L}_{\text{EFT}} = \mathcal{L}^{(\le4)} + \sum_{d=5}^{\mathcal{N}+4} \frac{1}{\Lambda^{d-4}} \sum_i C_i^{(d)} \mathcal{O}_i^{(d)} \,,
\end{equation}
where the so-called \emph{Wilson coefficients} $C_i^{(d)}$ parameterize the effective couplings.
They can be fitted to experimental measurements or matched to high-energy theories.

\section{Operator bases and redundancies}\label{sec:redundancies}
Determining an independent set of operators at a given mass dimension is not unique.
There are nontrivial relations between different operators, rendering a subset of them redundant.
The redundant operators are, however, indistinguishable in physical processes.
In practical applications, this leads to an over-determined set of Wilson coefficients so that no unique solution can be identified.
Consequently, it is necessary to avoid or eliminate these superficial operators and construct an independent \emph{operator basis} $\set{\mathcal{O}_i^{(d)}}$ for each mass dimension $d$.

Some \emph{redundancies} arise from the presence of (covariant) derivatives acting on the fields inside the operators.
Others are purely algebraic, related to the tensor structure of an operator.
Finally, we consider redundancies induced by the symmetry structure of the tensors linked to the occurrence of multiple fields transforming in equal representations under all symmetry groups inside an operator.
In the following, we display the crucial redundancies to consider in the operator basis construction.

\begin{description}
\item[\eoms]
Operators proportional to the \eom\ of some field $\Phi$ can be eliminated from the Lagrangian by an appropriate field redefinition in the path integral of the generating functional.
Consequently, operators $\mathcal{O}$ and $\mathcal{O}^\prime$, related by
\begin{equation}
\mathcal{O} \sim \mathcal{O}^\prime + \frac{\delta S}{\delta \Phi} \mathcal{O}^{\prime\prime} \,,
\end{equation}
where $S$ denotes the action, result in the same physical $S$-matrix elements.
It is, therefore, sufficient to only consider one of $\mathcal{O}$ and $\mathcal{O}^\prime$ in the on-shell basis.
\item[\ibp]
Operators proportional to total derivatives do not contribute to the perturbative action and can thus be eliminated from the Lagrangian.
Hence, operators $\mathcal{O}$ and $\mathcal{O}^\prime$, related by
\begin{equation}
\mathcal{O} \sim \mathcal{O}^\prime + \partial\mathcal{O}^{\prime\prime} \,,
\end{equation}
yield the same perturbative $S$-matrix elements, and only one of the operators should be considered as part of the \eft\ basis.
\item[Fierz identities]
Fierz identities relate products of spinor bilinears to linear combinations of products of different spinor bilinears.
It is, hence, possible to identify operators with apriori different spinor structures with each other.
Consider, for example, the identity
\begin{equation}\label{eq:fierz}
g_{\mu\nu} \sigma^\mu_{\alpha\dot\alpha} \sigma^\nu_{\beta\dot\beta} = 2 \epsilon_{\alpha\beta} \epsilon_{\dot\alpha\dot\beta} \,,
\end{equation}
where $\alpha,\beta$ and $\dot\alpha,\dot\beta$ denote spinor indices and $\mu,\nu$ denote Lorentz four-vector indices.
\Cref{eq:fierz} relates the direct contraction of the spinor indices on the \rhs\ to the contractions containing $\sigma$-matrices in between on the \lhs.
\item[Algebraic identities]
Algebraic identities relate the $\SU(N)$ tensor structures of different operators.
It is, therefore, possible to identify operators with different contractions of the internal symmetry indices.
Consider, for example, the identity
\begin{equation}
\epsilon^{ij}\epsilon^{kl} + \epsilon^{ik}\epsilon^{lj} + \epsilon^{il}\epsilon^{jk} = 0 \,,
\end{equation}
where $i,j,k,l$ denote fundamental indices of $\SU(2)$.
Since the same relation also holds for spinor indices instead, also called Schouten identity, it gives rise to additional Fierz identities as well.
\item[\rfrs]
Operators that contain the same field more than once can suffer from \rfrs.
On the one hand, operators $\mathcal{O}$ and $\mathcal{O}^\prime$ that only differ by a permutation $\pi_\Phi$ of the fields $\Phi$ inside the operator, i.e.,
\begin{equation}
\mathcal{O} \sim \pi_\Phi \circ \mathcal{O}^\prime \,,
\end{equation}
lead to the same physical processes.
This statement also remains valid if the field $\Phi$ appears in multiple generations, as the permutation of the generation indices can be absorbed into the Wilson coefficient tensor.
On the other hand, the explicit contraction of the fields can induce a symmetry on the generation indices, rendering some components of the Wilson coefficient tensors dependent.
\end{description}

\section{AutoEFT~--~An introduction}
Provided the field content and symmetries of a low-energy theory, \autoeft\ can be used to construct on-shell operator bases for, in principle, arbitrary mass dimensions automatically.
The resulting operator bases are free of redundancies by construction.
In the following, we briefly showcase the algorithm, the model files, and the output.
An exhaustive description of \autoeft\ can be found in \citere{Harlander:2023ozs}.

\begin{description}
\item[The algorithm]
A systematic algorithm, taking care of all the redundancies while simultaneously constructing explicit operator contractions, was proposed in \citeres{Li:2020gnx,Li:2020xlh}.
It is based on on-shell methods and various techniques from the representation theory of Lie groups and the symmetric group.
Contrary to existing efforts on the operator basis construction, the procedure does not generate a superset of operators, which is subsequently reduced by eliminating dependent operators using the redundancy relations introduced in \cref{sec:redundancies}.
Instead, the redundancies are avoided during the construction, so no superficial operator needs to be considered.
Notable exceptions to this are the \rfrs, which still need to be eliminated after obtaining a super basis of operators.
Due to the generality of the algorithm, it can be applied to various low-energy scenarios.
A schematic overview of the algorithm and its implementation in \autoeft\ is displayed in \cref{fig:algorithm}.

\item[The model file]
The model file encodes all information about the symmetries and field content of the low-energy theory and serves as input to \autoeft.
The model can contain arbitrary products of local and global $\SU(N)$ and $\U(1)$ symmetry groups.
The fields of the low-energy theory must be provided in irreducible representations under all symmetries.
It is, therefore, required to decompose any fields not transforming in irreducible representations beforehand.
For example, instead of providing a single bispinor, the user must define its irreducible components as two Weyl spinors.
The reasonably simple input format enables the user to compose custom model files and to construct the respective operator bases with minimal effort.
A variety of examples can be found in \citere{Harlander:2023ozs}.

\item[The output]
Due to the nature of the algorithm, the operators generated by \autoeft\ contain fields with spinor indices only.
For similar reasons, the fields carry only fundamental indices of the non-abelian $SU(N)$ groups.
Nevertheless, the operators can be translated to a more familiar notation by performing the substitutions provided in the appendix of \citere{Harlander:2023ozs}.
The resulting bases are provided in an electronic, human- and computer-readable output format.
In addition, it is possible to load operator bases with \autoeft, for example, to convert the operators to \LaTeX\ expressions, or for further manipulation of the explicitly contracted operators.
\end{description}

\definecolor{eftlightblue}{RGB}{142, 186, 229}
\definecolor{eftblack}{RGB}{0, 0, 0}
\definecolor{eftmaygreen}{RGB}{189, 205, 0}
\definecolor{eftorange}{RGB}{246, 168, 0}
\tikzstyle{eftinput} = [rectangle, rounded corners, thick, minimum width=1cm, minimum height=1cm, text centered, draw=eftblack, fill=eftmaygreen]
\tikzstyle{eftobject} = [rectangle, rounded corners, thick, minimum width=2cm, minimum height=1cm, text centered, draw=eftblack, fill=eftlightblue]
\tikzstyle{eftoutput} = [rectangle, rounded corners, thick, minimum width=1cm, minimum height=1cm, text centered, draw=eftblack, fill=eftorange]
\tikzstyle{arrow} = [thick, ->, >=stealth]
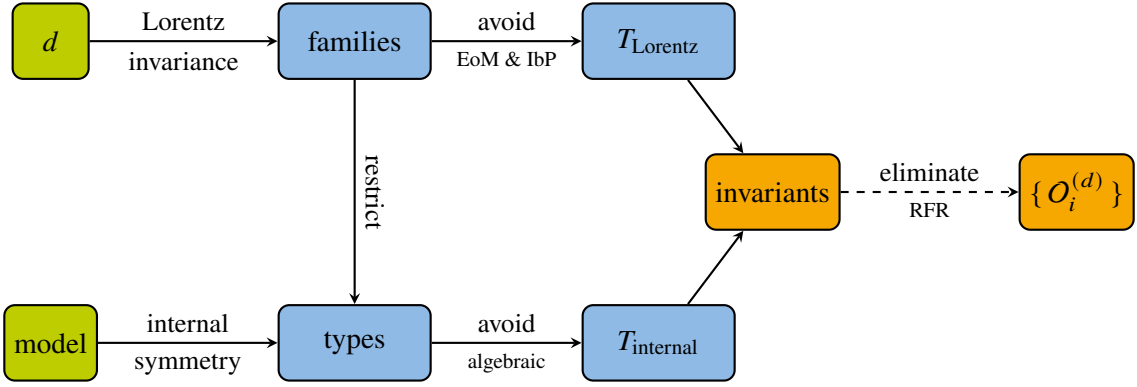
\begin{figure}
\centering
\begin{tikzpicture}[node distance=4cm]
  \node (fam) [eftobject] {families};
  \node (dim) [eftinput, left of=fam] {$d$};
  \node (TLor) [eftobject, right of=fam] {$T_{\text{Lorentz}}$};

  \node (type) [eftobject, below of=fam] {types};
  \node (mod) [eftinput, left of=type] {model};
  \node (TInt) [eftobject, right of=type] {$T_{\text{internal}}$};

  \node (invs) [eftoutput, below of=TLor, yshift=2cm, xshift=1.5cm] {invariants};
  \node (basis) [eftoutput, right of=invs] {$\set{\mathcal{O}_i^{(d)}}$};

  \draw [arrow] (dim) -- node[above] {\small Lorentz} node[below] {\small invariance} (fam);
  \draw [arrow] (mod) -- node[above] {\small internal} node[below] {\small symmetry} (type);
  \draw [arrow] (fam) -- node[sloped, above] {\small restrict} (type);
  \draw [arrow] (fam) -- node[above] {\small avoid} node[below] {\scriptsize EoM\ \& IbP} (TLor);
  \draw [arrow] (type) -- node[above] {\small avoid} node[below] {\scriptsize algebraic} (TInt);
  \draw [arrow] (TLor) -- (invs);
  \draw [arrow] (TInt) -- (invs);
  \draw [dashed,arrow] (invs) -- node[above] {\small eliminate} node[below] {\scriptsize RFR} (basis);
\end{tikzpicture}
\caption{
Schematic depiction of the algorithm implemented in \autoeft.
Green nodes denote the input required by \autoeft.
In this case, $d$ is the mass dimension.
Blue nodes label intermediate objects also available in \autoeft's output.
See \citere{Harlander:2023ozs} for a detailed description of their meaning.
The orange nodes denote the final bases of operators generated by \autoeft.
The first one, labeled \enquote{invariants}, is the super basis obtained by assuming that every field in the operator is distinct.
The second one is the (minimal) physical basis $\set{\mathcal{O}_i^{(d)}}$ without any \rfrs.
Solid arrows indicate that no elimination of superficial structures is required, as only independent sets of structures are generated.
The dashed arrow denotes a reduction of the basis by eliminating superficial operators.
}\label{fig:algorithm}
\end{figure}

\section{The standard model as effective field theory}
One of the most prominent \efts\ in high-energy physics is the so-called \smeft.
It can be used to parameterize the effects of heavy new physics at current energies in a model-independent manner.
It is obtained by identifying $\mathcal{L}^{(\le4)}=\mathcal{L}_{\text{SM}}$ in \cref{eq:eft}, where $\mathcal{L}_{\text{SM}}$ denotes the \sm\ Lagrangian.
The operators are thus composed of all \sm\ fields and must be invariant under the \sm\ symmetries $\LG(1,3) \times \SU(3) \times \SU(2) \times \U(1)$.
The repeated attempts needed to arrive at minimal \smeft\ bases, even at lower mass dimensions, demonstrate the complexity of constructing a complete and non-redundant set of operators.
For example, the first version of \citere{Grzadkowski:2010es} provided the operators of type $LQ^3$ divided into two terms.
Under the exchange of the first two quark fields, the two terms are symmetric and antisymmetric, respectively.
However, both of them contain the same mixed symmetry contribution so that the two terms are not independent~\cite{Abbott:1980zj,Alonso:2014zka}.

In the following, we use \autoeft\ to qualitatively examine this dependence.
Generating the operators for the mass dimension-six \smeft\ operator type $LQ^3$ with \autoeft, one finds three terms with definite permutation symmetry of the quark fields: $\mathcal{O}_{[3]}, \mathcal{O}_{[2,1]}, \mathcal{O}_{[1,1,1]}$.
Expressing the two terms $Q_{qqq}^{(1)}$ and $Q_{qqq}^{(3)}$ of \citere{Grzadkowski:2010es} in terms of the \autoeft\ basis yields
\begin{equation}
Q_{qqq}^{(1)} = -\frac13 \mathcal{O}_{[3]} + \frac13 \mathcal{O}_{[2,1]}
\qquad\text{and}\qquad
Q_{qqq}^{(3)} = -\frac13 \mathcal{O}_{[2,1]} - \mathcal{O}_{[1,1,1]} -\frac23 \mathcal{O}_{[2,1]}^{(23)} \,,
\end{equation}
where $\mathcal{O}_{[2,1]}^{(23)}$ is obtained from $\mathcal{O}_{[2,1]}$ by interchanging the second and third quark fields.
At this point, the dependence is already evident since both terms contain the mixed symmetry contribution from $\mathcal{O}_{[2,1]}$.
In addition, one finds that neither $Q_{qqq}^{(1)}$ nor $Q_{qqq}^{(3)}$ contain all of $\mathcal{O}_{[3]}, \mathcal{O}_{[2,1]}, \mathcal{O}_{[1,1,1]}$ and thus do not span the entire space.
However, by changing basis to
\begin{equation}
Q_{qqq} = -\frac12 (Q_{qqq}^{(1)} + Q_{qqq}^{(3)})
\qquad\text{and}\qquad
\tilde{Q}_{qqq} = -\frac12 (Q_{qqq}^{(1)} - Q_{qqq}^{(3)}) \,,
\end{equation}
one finds that $Q_{qqq}$ contains each term exactly once.
In addition, one can identify $Q_{qqq}^{(12)} = \tilde{Q}_{qqq}$, rendering $\tilde{Q}_{qqq}$ redundant.
It is thus sufficient to consider only the term $Q_{qqq}$ in the operator basis, as was done in later versions of \citere{Grzadkowski:2010es}.

\section{EFT bases}
We have used \autoeft\ to derive for the first time the on-shell \smeft\ operator bases for mass dimensions 10, 11, and 12~\cite{Harlander:2023psl}.
In addition, by exploiting the generality of the algorithm, we constructed operators including gravitational interactions up to mass dimension 12.
We provide the bases generated by \autoeft\ in a public repository~\cite{eft_bases_repository}.
It can be used to examine different operator bases, either by hand or by loading them into \autoeft.
Currently, we provide the operator bases for \smeft, its extension by gravitational interactions, and its restriction to minimal flavor-violating scenarios.
We intend to extend the set of bases to other models in the future.
In addition, users of \autoeft\ have the opportunity to submit operator bases to the repository.

\section{Summary}
\efts\ have become an indispensable tool in the search for heavy new physics.
We have introduced the application \autoeft, developed for the construction of on-shell operator bases with minimal effort and for various low-energy scenarios.
Besides the operator construction, \autoeft\ can be used to examine the relations between different operator bases.
It provides a foundation for future \eft\ frameworks, and we intend to extend its capabilities in various respects.

\acknowledgments{%
I would like to thank Robert Harlander, Tim Kempkens, Jakob Linder, and Maximilian Rzehak for their inspiring collaboration.
This research was supported by the Deutsche Forschungsgemeinschaft (\abbrev{DFG}, German Research Foundation) under grant \gepris{400140256}~–-~\textit{GRK~2497: The physics of the heaviest particles at the LHC}, and grant \gepris{396021762}~-–~\textit{TRR~257: P3H~-–~Particle Physics Phenomenology after the Higgs Discovery}.%
}

\bibliographystyle{JHEP}
\bibliography{proceeding}

\end{document}